\documentstyle[prd,aps,preprint]{revtex}

\catcode`@=12
\begin{document}
\vspace{0.5in}
\oddsidemargin -.375in  
\newcount\sectionnumber 
\sectionnumber=0 
\def\be{\begin{equation}} 
\def\ee{\end{equation}}
\thispagestyle{empty}
\begin{flushright} BIHEP-TH-96-31 \\
BROWN-HET-1064 \\
November 1996\\
\end{flushright}
\vspace {.5in}
\begin{center}
{\Large \bf{Cosmic Strings In Theories With Dynamical Symmetry Breaking\\ }}

\vspace{.5in}
{\rm R. Brandenberger$^{a}$, T. Huang$^{b}$, K. Yang$^{b}$
 and X. Zhang$^{b}$\\} 
\vspace{.3in}

{$^a$
\it Department of Physics, 
 Brown University, Providence RI. 02912, USA.
\\ }

{$^b$
\it Institute of High Energy Physics, Academia Sinica, P.O. Box 918(4), \\
 \it Beijing 100039, P.R. China.
     \\}
\vskip .8in
\end{center}  
\begin{abstract}
We study the classical cosmic string solution in a theory with dynamical $U(1)$
symmetry breaking. We calculate the energy per unit length of the
string and compare it to that obtained in a model with a fundamental Higgs.
We find that the predictions of the standard Abelian Higgs model are quite 
stable towards the addition of higher dimension operators expected in the 
effective Lagrangian for the order parameter in a model with dynamical 
symmetry breaking, at least for small coupling constant $\lambda$.
\end {abstract}
\newpage
\baselineskip 24pt

The role of cosmic strings for structure formation of the Universe
is determined
by the energy per unit length $\mu$. This energy determines the amplitude of the
spectrum of density perturbations and of microwave anisotropies.
In models, where a gauged $U(1)$ is broken by a fundamental Higgs $\Phi$
with potential $V( \Phi ) = \lambda {( \Phi^\dagger \Phi - \eta^2 )}^2$,
$\mu$ and $\eta$ are related via $\mu \simeq \eta^2$.
If cosmic strings are indeed responsible for seeding the large-scale 
structure of the Universe, then to explain the observed data, 
a value of $\eta \simeq 10^{16}$GeV is required.  
The interesting question now is: does this result change if 
the $U(1)$ symmetry breaking is caused not by a fundamental
Higgs field, but by an order parameter similar to the
quark pair condensate in QCD? 
The answer to this question is very
relevant to the robustness of the predictions of the cosmic string model.

Dynamical symmetry breaking at the electroweak scale
has been widely
discussed. Models for the breakdown of $SU(2)_L \times U(1)_Y$ have been 
proposed, such as Composite Higgs models\cite{Georgi}, top 
condensation\cite{hill}, and Technicolor\cite{tc} and many others.
Dynamical breakdown of the Peccei-Quinn symmetry in the composite Axion
model\cite{kim} and in the supersymmetric preon models\cite{pati} serve
as examples of dynamical symmetry breaking above the electroweak scale.

Experimentally, with an 
accelerator with energy reaching the symmetry breaking scale, one could 
possibly tell the difference
between the two scenarios of symmetry breaking.
An example is provided by the precise measurement of the properties of the 
$Z^0$ particle at LEP I.
The data on S. T. U\cite{peskin} and $R_b$\cite{chivukula} 
disfavor the conventional technicolor
models. However, it will be much difficult to obtain direct experimental
information on the dynamics of the phase transition if
the symmetry breaking scale is far above the energies provided by the current
accelarators, such as is the case for
the breaking of the $U(1)$ which generates strings of relevance for 
structure formation. In this case, a different approach must be taken. 
In this paper we will examine the effects of the two scenarios of symmetry 
breaking on the properties of the topological soliton. Specifically,
we will calculate the energy per unit length $\mu$ of a cosmic string in
a theory with dynamical $U(1)$ symmetry breaking and compare the result to that
obtained with a fundamental Higgs scalar. 

We begin with a discussion of the effective
Lagrangian in a theory with dynamical $U(1)$ symmetry
breaking.  In the Higgs model one introduces
a fundamental complex scalar field $\Phi$,
which transforms non-trivially under $U(1)$.
The lagrangian required by renormalizability has the following form:
\begin{equation}
{\cal L} = -\frac{1}{4} F_{\mu\nu} F^{\mu\nu}
+ (D_\mu \Phi)^{\dagger} (D^\mu \Phi)
- \lambda (\Phi^{\dagger} \Phi - \eta^2 )^2,
\   \end{equation}
where
$
F_{\mu\nu}=\partial_\mu A_\nu-\partial_\nu A_\mu $
is the U(1) field strength tensor, 
$D_{\mu} \Phi = \Bigl(\partial_{\mu}+ie A_{\mu} \Bigr)\Phi$ is the covariant
 derivative of $\Phi$,
and the coupling constant is denoted by $e$, which we take to be
of the order of the electric charge in our numerical calculation.
However, in a theory with dynamical symmetry breaking, the order parameter,
still denoted by $\Phi$, will not be fundamental. Associated with the 
structure of the $\Phi$ field, there is an energy scale above which the 
effective lagrangian breaks down.
In general, the effective lagrangian
will include higher dimension operators in addition to those present in eq.(1),
\begin{equation}
{\cal L}^{\rm eff} = {\cal L} + \Delta {\cal L}, \
\end{equation}
where $\Delta{\cal L}$ includes all of the higher dimensional operators, which
are $U(1)$ gauge invariant and consist of the gauge field $A_\mu$ and scalar 
$\Phi$.
Given such an effective lagrangian ${\cal L}^{\rm eff}$,
we are now to look for the static vortex solution
and calculate its energy per unit length. To begin with, let us consider first
the dimension-six operators. Specifically, they are given by
\footnote{ 
We impose renormalization conditions so that the kinetic energy of the gauge and
Higgs fields in the broken phase will not be renormalized by the
 higher dimension operators. So we use, for instance, ${ ( D_\mu \Phi )}^\dagger
(D^{\mu}\Phi)(\Phi^\dagger \Phi -\eta^2 )$ instead of
${ ( D_\mu \Phi ) }^\dagger ( D^\mu \Phi )(\Phi^\dagger \Phi)$.
}
\begin{equation}
\Delta{\cal L}_{6} = \frac{C_1}{\Lambda^2}(\Phi^{\dagger} \Phi- \eta^2 )^3
                +\frac{C_2}{\Lambda^2}
                 (D_\mu \Phi)^{\dagger} (D^\mu \Phi)
                 (\Phi^{\dagger} \Phi - \eta^2 )
                +\frac{C_3}{\Lambda^2}
                (-\frac{1}{4} F_{\mu\nu} F^{\mu\nu})
                 (\Phi^{\dagger} \Phi - \eta^2 ),
\  \end{equation}
where the coefficients $C_i$ $(i=1,2,3)$ give the strength of the 
contribution of
the new physics to the effective theory, and where
$\Lambda$ is the cutoff. In a theory with dynamical symmetry breaking, 
$\Lambda$ is of order of the vacuum
expectation value $\eta$. In our numerical calculations, we take $\Lambda
= 5~ \eta$. And
one expects in general that $C_i \sim O(1)$ in dynamical symmetry breaking
theories.

Static vortex solutions of the Abelian Higgs model were considered in a
 classic paper by Nielsen and Olesen\cite{string} 
(see also Ref. \cite{vortex} for a discussion of vortex solutions in 
superconductors and Ref. \cite{CSreviews} for a general overview), 
 \begin{eqnarray}
   A_0 & = & 0,   
\nonumber \\
  {\bf A} & = & \frac{1}{er}[1-F(r)]\hat \theta,
\nonumber \\
  \Phi & = & \rho(r)e^{i\theta}
\ , \end{eqnarray}
where $F(r)$ and $\rho(r)$ are radial functions.
No exact solutions of the resulting Euler-Lagrange equations are known, 
only approximate solutions for large and small values of $r$ \cite{string} 
(see also Ref. \cite{periv} for a recent careful analysis).

As is well known\cite{CSreviews}, for the Abelian Higgs model the energy per 
unit length, $\mu$, of the vortex depends only very weakly on the Higgs self 
coupling constant $\lambda$. Heuristically, this can be seen as follows. 
We expect the gauge and Higgs field to be,
\begin{eqnarray}
      F(r) & = & e^{- r/{r_c}}, 
\nonumber \\
      \rho(r) & = & \eta (1 - e^{-r/w}), 
\, \end{eqnarray}
where $r_c$ and $w$ are two variational parameters. 
The angular gradient and potential energies (per unit length) $\mu_{\rm{ang}}$
and $\mu_{\rm{pot}}$, respectively, then become
\begin{eqnarray}
\mu_{\rm{ang}} & = & 2 \pi \int_0^{\infty} dr r 
          [(D_{\mu} \Phi)^\dagger(D^\mu \Phi)]_{\rm{ang}}
\simeq 2 \pi \eta^2 \int_0^{r_c} dr r ({1 \over r})^2 (1 - e^{- r/w})^2 
\nonumber \\
& \simeq & 2 \pi \eta^2 \int_0^w dr {1 \over r} ({r \over w})^2
+ 2 \pi \eta^2 \int_w^{r_c} dr {1/r} \nonumber \\
& = & \pi \eta^2 + 2 \pi \eta^2 {\rm ln}({{r_c} \over w}) \, ,
\end{eqnarray}
and
\begin{equation}
\mu_{\rm{pot}} = 2 \pi \lambda \int_0^{\infty} dr r 
(\Phi^\dagger\Phi - \eta^2)^2
\simeq \pi \lambda \eta^4 w^2 \, .
\end{equation}
Minimizing the sum of these two energies with respect to the Higgs width 
$w$ yields  
\begin{equation}
w \simeq \lambda^{-1/2} \eta^{-1} \, ,
\end{equation}
from which it follows that the total energy per unit length depends only 
logarithmically on $\lambda$, increasing as $\lambda$ grows.

Higher dimension terms in the effective Lagrangian could very well change the 
powers of $w$ appearing in the expressions for gradient and potential energies,
thus resulting in a energy per unit length $\mu$ which depends more sensitively 
on $\lambda$, and in a different relation between $\mu$ and $\eta^2$. 
In the following we shall demonstrate that the results derived from the 
Abelian Higgs model are, at least for small $\lambda$, quite stable.

The energy functional in our theory including dimension-six operators is 
given by
\begin{equation}
\mu = \mu_{0 } + \mu_{6} 
\ , 
\end{equation}
where
\begin{equation}
 \mu_{0}  =  2\pi \int^{\infty}_0 dr r
         \Bigl[ \frac{1}{2}(\frac{1}{er})^2 F'^2 
         + \frac{F^2\rho^2}{r^2}
         +\rho'^2+\lambda(\rho^2-\eta^2)^2 \Bigl], 
\ \end{equation}
and
\begin{eqnarray}
\mu_{6} = 2\pi\int^{\infty}_0 dr r\Bigl[
            \frac{C_1}{\Lambda^2}(\rho^2-\eta^2)^3
         &+& \frac{C_2}{\Lambda^2}
           [(\frac{F^2\rho^2}{r^2}
         + \rho'^2)(\rho^2-\eta^2)] 
\nonumber \\
   & + & \frac{C_3}{\Lambda^2}
          [\frac{1}{2}(\frac{1}{er})^2 F'^2](\rho^2-\eta^2) \Bigl].
\   \end{eqnarray}
It is straightforward to derive the Euler-Lagrange equations for this energy
functional. They are given by
\begin{eqnarray}
     [1+\frac{C_3}{\Lambda^2}(\rho^2-\eta^2)]F''  
     &-&[\frac{1}{r}-\frac{C_3}{\Lambda^2}(2\rho\rho'
      -\frac{\rho^2-\eta^2}{r})]F' 
\nonumber \\
   &-&2e^2[\rho^2+\frac{C_2}{\Lambda^2}\rho^2(\rho^2-\eta^2)]F=0
\ , \end{eqnarray}
and
\begin{eqnarray}
     [1+\frac{C_2}{\Lambda^2}(\rho^2-\eta^2)]\rho''  
      &+&\frac{C_2}{\Lambda^2}\rho\rho'^2
      +[\frac{1}{r}+\frac{C_2}{\Lambda^2 r}
       (\rho^2-\eta^2)]\rho'
       -\frac{3C_1}{\Lambda^2}\rho(\rho^2-\eta^2)^2
\nonumber \\
        &-&2(\lambda+\frac{C_2}{\Lambda^2 r^2}F^2)\rho^3
         +(2\lambda \eta^2+\frac{C_2\eta^2}{\Lambda^2 r^2}F^2
         -\frac{F^2}{r^2}-\frac{C_3}{2e^2\Lambda^2 r^2}F'^2)\rho=0
\ . \end{eqnarray}

The boundary conditions for
$F$ and $\rho$ are given by
\begin{equation}
    F(0)=1,  \qquad  \rho(0)=0  
\ , \end{equation}
and
\begin{equation}
    F(\infty)=0,  \qquad \rho(\infty)=\eta  
\ . \end{equation}
Note that these are the same boundary conditions as those in the absence
of the higher dimension operators. We numerically integrate the equations
by minimizing $\mu$
for a given value of the parameter $\lambda$. 
In Fig. 1 we plot the values of $\mu$ obtained in the present case together 
with the corresponding result in the fundamental Higgs model 
({\it i.e.} in the absence of the higher dimensional operators). One can
see that for small $\lambda$, the new physics from
dynamical symmetry breaking gives only small correction to the
energy per unit length $\mu_0$ of less than about
$50 \%$.
This could be understood by using the simple ansatz given by eq. (5). Including
the dimension-six operators, we minimize the total energy and obtain that 
$w \sim 1/(\eta\sqrt{\lambda+0.01})$ and that 
$\mu_6 \sim 0.005\eta^4 w^2 < 0.5\eta^2$.   

As $\lambda$ gets large, the effects of the new physics on $\mu_0$
becomes negligible. This is because in the limit of $\lambda \rightarrow
\infty, \Phi \rightarrow \eta$,
so,
${ (\Phi^\dagger \Phi - \eta^2 )} \rightarrow 0$. In summary, for the whole
range of the parameter $\lambda$, the correction to $\mu$ due to 
dimension-six operator is less than about $50 \% $.

$\lambda$. 

Let us now consider the dimension-eight operators. For simplicity, we pick
one of the several invariant operators for a detailed calculation,
\begin{equation}
{\cal L}_{8} = \frac{C_4}{\Lambda^4}
                    \Bigl[ (D_\mu \Phi)^{\dagger}(D^\mu \Phi) \Bigl]^2.
\  \end{equation}
Focusing only on the contribution of this dimension-eight operator,
the energy per unit length of the cosmic string becomes
\begin{equation}
\mu = \mu_0 + \mu_{8}
\end{equation}
where
\begin{equation}
\mu_{8} = 2\pi\int_{0}^{\infty}dr r \Bigr[
          \frac{C_4}{\Lambda^4}
          (\frac{F^2\rho^2}{r^2}+\rho'^2)^2 \Bigr]
\ . \end{equation}
It is straightforward to derive the Euler-Lagrange equations for the 
above energy  functional. They are given by 
\begin{equation}
     F''-\frac{1}{r}F'-2e^2F\rho^2[1+\frac{2C_4}{\Lambda^4}
         (\frac{F^2\rho^2}{r^2}+\rho'^2)]=0
\ , \end{equation}
and
\begin{eqnarray}
     [1+\frac{2C_4}{\Lambda^4}(\frac{F^2\rho^2}{r^2}+3\rho'^2)]\rho''  
      +\frac{2C_4}{\Lambda^4 r}\rho'^3
      &+&\frac{2C_4F^2\rho}{\Lambda^4 r^2}\rho'^2
          +(\frac{1}{r}+\frac{4C_4FF'\rho^2}{\Lambda^4 r^2}
       -\frac{2C_4F^2\rho^2}{\Lambda^4 r^3})\rho'
\nonumber \\
          &-&(2\lambda+\frac{2C_4F^4}{\Lambda^4 r^4})\rho^3
           +(2\lambda\eta^2-\frac{F^2}{r^2})\rho=0
\ . \end{eqnarray}
Following the procedure above for the dimension-six operators, we obtain
the energy per unit length for the cosmic string with and without the
presence of the dimension-eight operator. The results are plotted in Fig.2.
One can see from the figure that for small $\lambda$, the correction of 
$\mu_0$ due to the dimension-eight operator is negligible. 
The two energies are practically indistinguishable. 
However, the difference becomes important for large $\lambda$. 
For instance, for $\lambda \sim
10^{5}$, the value of $\mu$ is greater than $\mu_0$ by a
factor 4. 
To understand it
we use again the ansatz in eq. (5) and obtain that
\begin{equation} 
 \mu_8 \sim \frac{\eta^6}{\Lambda^4} \lambda .
\  \end{equation}
This makes the energy $\mu$ depend on $\lambda$ linearly,
which is in contrast to the logrithmic dependence of the $\mu_0$ on 
$\lambda$ in the Higgs model shown in eqs.(6-8).

Let us now compare the effect on $\mu$ due to the dimension-eight
operator with that due to the dimension-six operators. For small $\lambda$,
the dimension-six operators make a larger contribution to the $\mu$
than the dimension-eight operator, since it is less suppressed by $\Lambda^2$.
However, for large $\lambda$, the dimension-eight operator is more important.
To understand it
let us consider
the limit $\lambda \rightarrow \infty$.
In this limit,
$\Phi \rightarrow \eta$, and the Lagrangian becomes 
 non-linearly realized, ${\cal L}^{\rm eff}_{\rm non-linear}$. In
${\cal L}^{\rm eff}_{\rm non-linear}$, 
the terms corresponding to the dimension-six
operators in eq. (3) vanish. The dimension-eight
operator in eq. (16) is a higher derivative term contributing to the
tension energy of the static vortex. This term starts to dominate over the 
canonical dimension-four kinetic term when the width of the vortex decreases, 
i.e for large values of $\lambda$. This effects the width of the vortex, 
as shown in Fig. 3. 
This explains why the dimension-eight operator is more
important than the dimension-six operators for large $\lambda$. Furthermore,
since the dimension-eight operator corresponds to the first higher order
term in the ${\cal L}^{\rm eff}_{\rm non-linear}$, 
the operators with higher dimension  
than eight will become less important than the one with dimenion-eight.

In summary, we have numerically studied the effects of
dynamical $U(1)$ symmetry breaking  on the properties of the static vortex 
solution. Even though the topological argument garantees the existence of the
cosmic string, its properties, in particular the energy per unit length
$\mu$, depend on the symmetry breaking mechanism. For small
values of the quartic self coupling constant $\lambda$, the two energy $\mu$
and $\mu_0$ are not very different, so the formula,
 $\mu \simeq \eta^2$ is still approximately valid\footnote{
Figure 1 indicates that $\mu \sim O( \eta^2)$ for a very large
parameter space in $\lambda$.}. For large values of $\lambda$, which
is also preferred in the scenario of the dynamical symmetry breaking,
the energy per unit length $\mu$ will much higher than the value $\mu_0$ 
obtained in a theory with a fundamental Higgs, and $\mu >> \eta^2$. 
Consequently, $U(1)$ dynamical symmetry breaking at a scale smaller
than $10^{16}$GeV may be consistent with the hypothesis that cosmic strings 
form the seeds for structure in the Universe.  

{\bf ACKNOWLEDGMENTS}

We would like to thank C. Huang for discussions. This work was supported in
part by the National Science Foundation of China (T. Huang, K. Yang 
and X. Zhang) and by the U.S. Department of Energy under contract DE-FG0291ER40688, Task A (R. Brandenberger).


\section{Figure Captions}

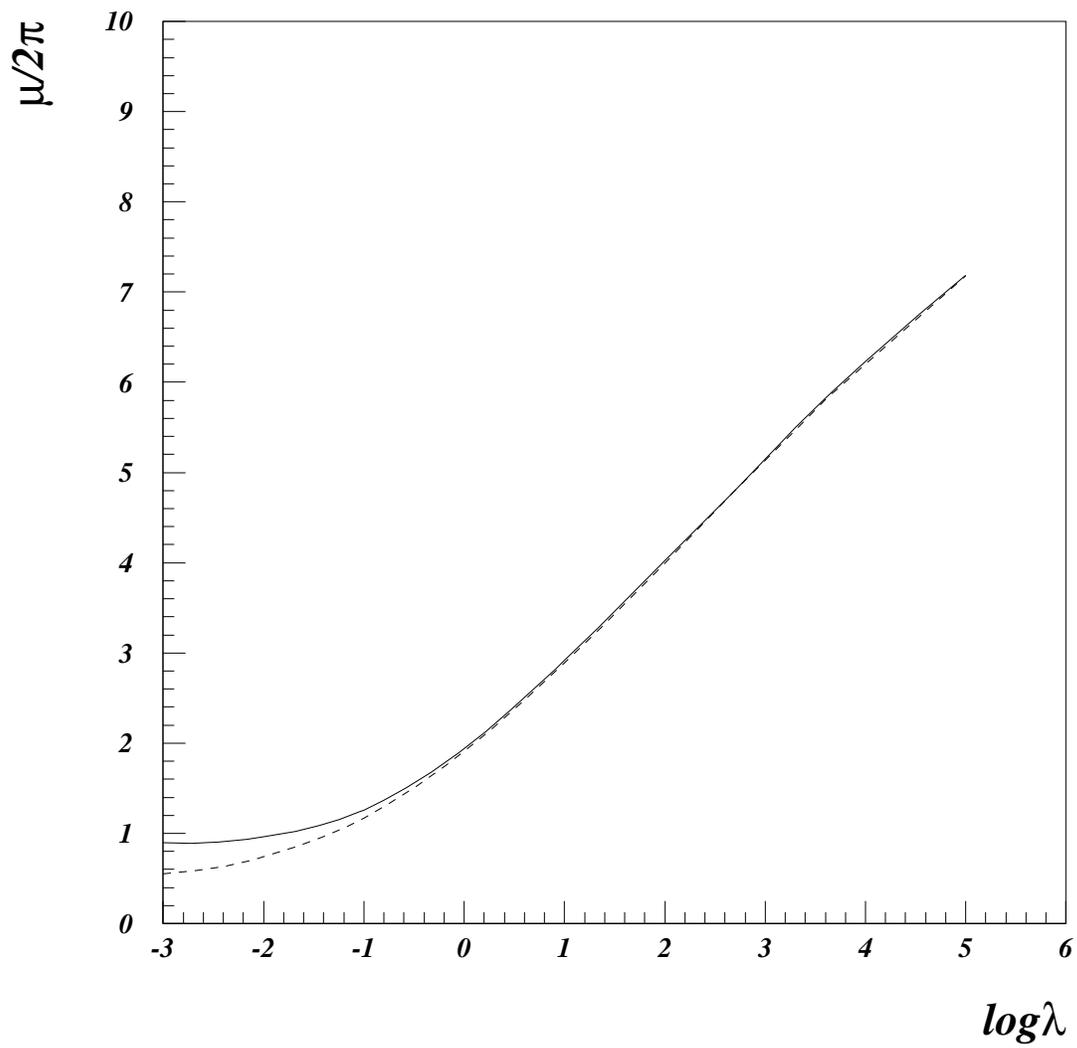
\noindent{\bf Fig. 1} 
The figure shows the energy per unit length in units of 
$\eta^2$ vs $\log (\lambda)$.
The solid line    
is for $\mu$, dashed one for $\mu_0$. In the numerical calculation, we take 
$C_1= C_2 =C_3 = -1$.

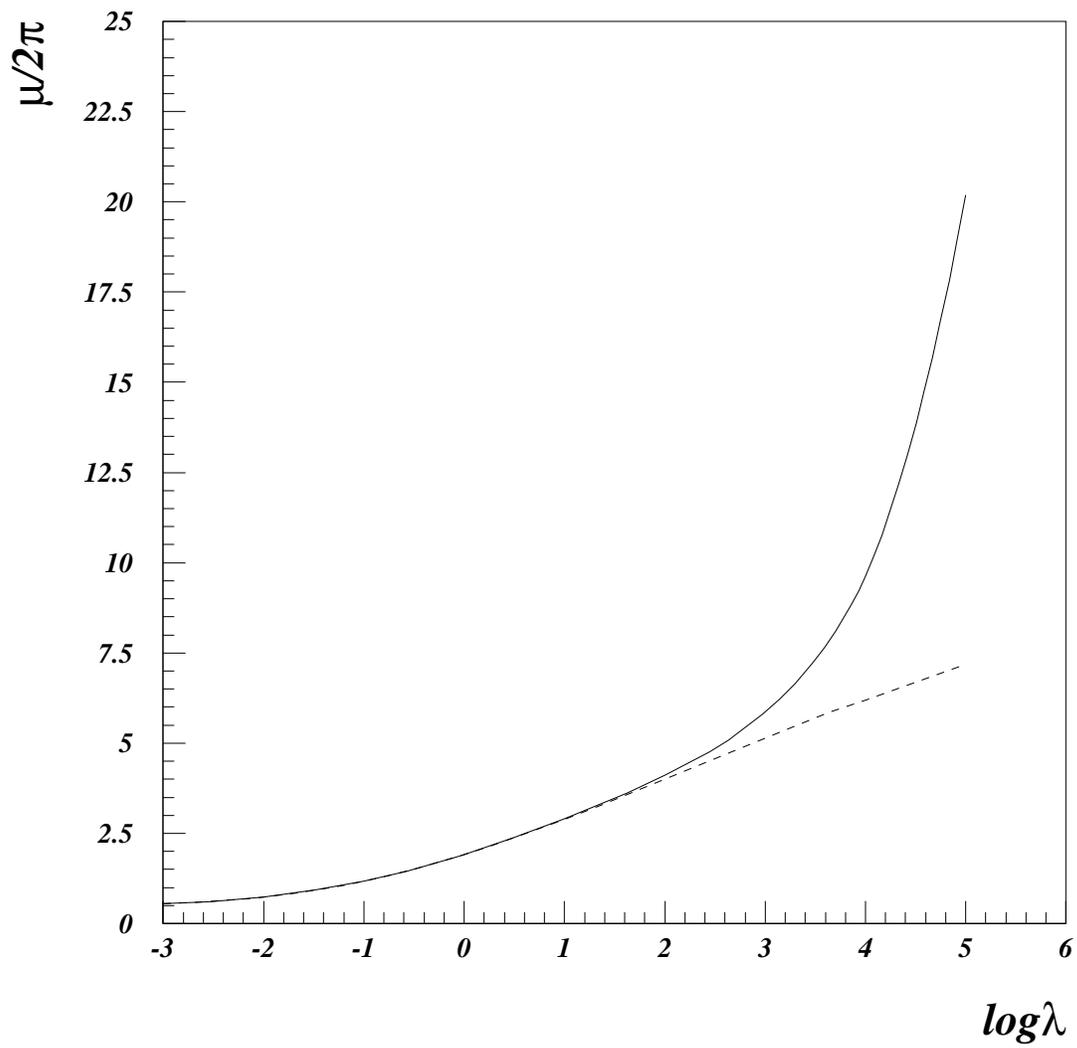
\noindent{\bf Fig. 2}
The energy per unit length
in unit of $\eta^2$ vs $\log (\lambda )$. The solid line is for
$\mu$, and the dashed one for $\mu_0$. In the numerical calculation, we have
taken $C_4 = 1 $.  

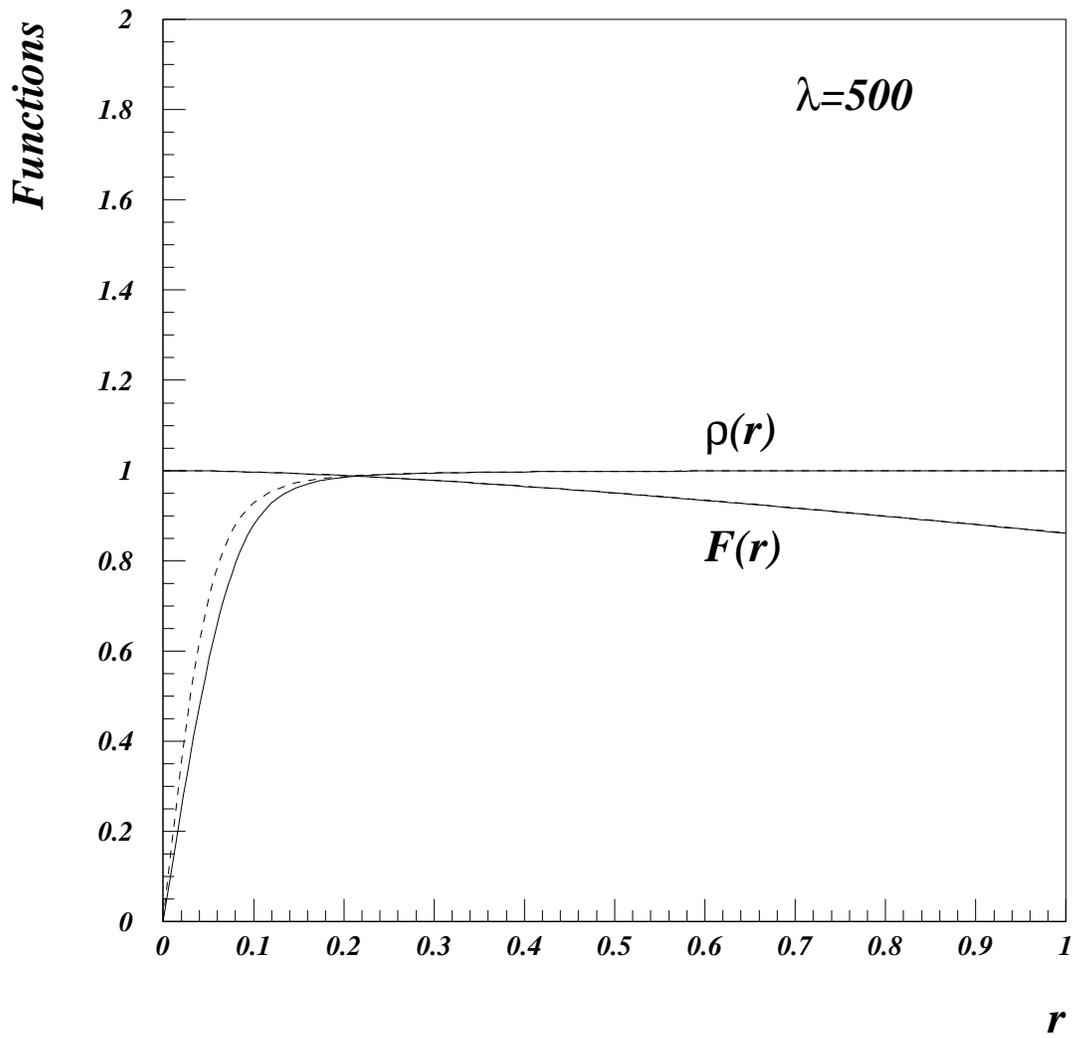
\noindent{\bf Fig.3}  
The radial functions $F(r)$, $\rho(r)$ of the
string. The solid curves correspond to the model
with the dimension-eight operator and the dashed ones to 
the fundamental Higgs model. The functions $F(r)$ for the two models are 
indistinguishable.

\end{document}